\documentclass[aps,prl,twocolumn,floats,showpacs]{revtex4}
\usepackage{graphicx}
\begin{document}
\title{High-order current correlation functions in  Kondo systems.}
\author{A. Golub}
%and $^{1,2}$ }
\affiliation{ Department of Physics, Ben-Gurion University of the
Negev, Beer-Sheva, Israel \\   }
\begin{abstract}
We examine the statistics of current fluctuations  in a junction
with a quantum dot described by Kondo Hamiltonian. With the help
of modified Keldish technique we calculate the third current
cumulant. As a function of ratio $v=eV/T_{K}$ the 3rd cumulant was
obtained for three different regimes: Fermi liquid regime ($v<1$),
crossover interval ($v\geq1$) and RG limit ($v>>1$). Unlike the
case of noninteracting dot, 3rd cumulant shows strong non-linear
voltage dependence. Only in the asymptotical limit
$v\rightarrow\infty$ the linear dependence on $V$ is recovered.
\end{abstract}
\maketitle
 {\it Motivation}: The direct electron transport is an important
  tool to study small
 junctions and quantum dots.
 The investigation of the current correlation functions helps to get
 additional information
 about physical properties of such systems. During last years the
 measurements only of the second cumulant of
 fluctuating current (shot noise) have resulted in a
 remarkable theoretical and
 experimental progress (see review article \cite{but}).
 Interest in the third
and higher moments has occurred, first, because its
characteristics
 differ significantly from that of the second moment.
In particular, as it was predicted by Levitov and
 Reznikov \cite{reznikov}, in non-interacting systems the third moment is
insensitive to the sample's thermal noise, yet is more sensitive
to the environment. Second, measurements of the higher moments may
provide a new tool for studying conduction physics, complementary
to the second moment.
 Experimentally  it is quite nontrivial
 to extract cumulants higher then the second.
 Recently, however, the first experimental study
 of the third current
cumulant in mesoscopic tunnel junctions was reported
\cite{reulet}.

The theory of full counting statistic \cite{levitov} is a
theoretical framework which is used to analyze statistics of
charge transfer in experiment and to calculate the higher order
cumulants \cite{nazarov,nagaev}. However, when the interactions
are included, which is the subject of the present paper, it is
necessary to go beyond the theory of full counting statistics.
Before starting such an analysis  for Kondo type quantum dots we
note that recently in a number of  works  third current cumulant
in interacting mesoscopic systems was calculated
\cite{gefen,zaikin}. In these works the 3rd current cumulat was
defined as a time ordered product of three current operators on
standard Keldish 2-time contour. As a result this correlation
function is equal to a sum of partially time ordered products on
the usual one line time contour. Thus we can apply Feynman
perturbation theory to evaluate this correlation function,
however, we are unable for interacting systems to compute non-time
ordered 3rd cumulant. It is important to resolve this discrepancy.
Here with the help of a modified Keldish technique which uses a
multiple time contour (see Fig.1) \cite{kogan} we directly provide
such a derivation of non-time ordered 3rd cumulat for the Kondo
problem.

 {\it Hamiltonian. Third cumulant}: We start with the Kondo Hamiltonian for quantum
 dot in the junction $H=H_{L}+H_{R}+H_{J}$ where
\begin{eqnarray}
H_{J}&=&\sum_{\alpha\alpha'\sigma\sigma'}J_{\alpha
\alpha'}c^{\dagger}_{\alpha\sigma}(0)(\frac{1}{4}\delta_{\sigma,\sigma'}+
\vec{S}\vec{\emph{s}}_{\sigma\sigma'})c_{\alpha'\sigma'}(0)
\label{H2}
\end{eqnarray}
\begin{figure}[ht]
\begin{center}
\includegraphics [width=0.4\textwidth ]{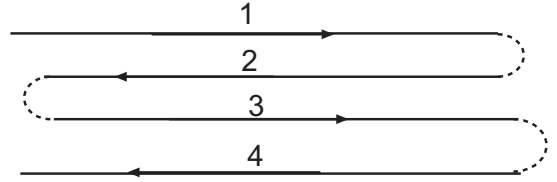}
\caption {\label{fig1} The four-axis time contour }
\end{center}
\end{figure}
The first two terms correspond to non-interacting electrons in the
two leads
$H_{L(R)}=\sum_{k,\sigma}\xi_{L(R)k}c^{\dagger}_{L(R)\sigma,k}c_{L(R)\sigma,k}$,
 where $c_{\alpha\sigma,k}$, $ \xi_{\alpha k}$
are the electron field operator  and the electron energy of a
lead. Index $\alpha=L,R$ indicates left (right) lead. We assume
that the leads are dc-biased by applied voltage $V$. Here
$\emph{s}$ is the one half spin matrix which acts on spin index of
electron operators and $S$ is the spin operator of the dot. The
potential scattering is represented by the first term in the
brackets. The bare coupling constants in (\ref{H2}) can be
obtained by Schrieffer-Wolff transformation from the parent
Anderson Hamiltonian
\begin{eqnarray}
\hat{H}&=&H_{L}+H_{R} + \!\sum_{k,\sigma,\alpha}(v_\alpha
c^{\dagger}_{\alpha \sigma,k}d^{\phantom{\dagger}}_{\sigma}+ {\rm
H.c.}
)\nonumber\\
&+&\sum_\sigma \epsilon d^{\dagger}_{\sigma}
d^{\phantom{\dagger}}_{\sigma}+
Ud^{\dagger}_{\uparrow}d^{\phantom{\dagger}}_{\uparrow}
d^{\dagger}_{\downarrow}d^{\phantom{\dagger}}_{\downarrow}\
\label{HA}
\end{eqnarray}
and are related to the parameters of this Hamiltonian
\cite{glazman} as ${ J}_{\alpha\alpha'}=
\sqrt{\Gamma_\alpha\Gamma_{\alpha'}}/(\pi\nu\tilde{|\epsilon|})$
and $\quad \tilde{\epsilon}\equiv(U-|\epsilon|)\epsilon/U $. Here
$U$ is the repulsive Hubbard coupling, $\epsilon$ denotes bare
level energy of the dot and $\nu$ is the density of states in a
lead. We also introduce the widths $\Gamma_{L,R}=2\pi \nu
|v_{L,R}|^2\ $ which are expressed in terms of tunnelling matrix
elements and density of electron states \cite{ygal}. The current
operator has a form
\begin{eqnarray}
I&=&\frac{ie}{\hbar}[J_{LR}
\sum_{\sigma\sigma'}c^{\dagger}_{L\sigma}(0)(\frac{1}{4}\delta_{\sigma,\sigma'}+
\vec{S}\vec{\emph{s}}_{\sigma\sigma'})c_{R\sigma'}(0)-H.C.]\nonumber
\end{eqnarray}
 We define the 3rd momentum noise as a symmetric
combination of non-time ordered correlation function of three
currents. In a stationary situation this function can be written
as
\begin{eqnarray}
{\cal S}_{3}(t_1 -t_2,t_2 -t_3
)&=&\frac{1}{6}\sum_{P(\alpha\beta\gamma)}<I(t_\alpha
)I(t_\beta)I(t_\gamma )>\label{st}
\end{eqnarray}
where $P(\alpha\beta\gamma)$ is the permutation of 1,2,3.
 The Fourier-transformed value of (\ref{st})
${\cal S}_{3}(\omega_{1} ,\omega_2 )$ is a function of two energy
variables.
 Below we consider the zero frequency limit
 ${\cal S}_{3}(\omega_1=0
,\omega_2 =0)\equiv {\cal S}_{3}$ and  introduce the third order
cumulant $S_{3}$, that is, the irreducible part of correlation
function (\ref{st}). This cumulant yields the equation
$S_{3}={\cal S}_3 -3\bar{I}S_2+2\bar{I}^3$. Here $\bar{I}$ is the
averaged current and $S_2$ stands for the pair current correlation
function (shot noise). In the Kondo regime this function was
recently calculated \cite{us}.

{\it Perturbation theory. RG}: To apply perturbation theory to a
product of non-time ordered Heisenberg operators like $S_{3}$ we
need to order these operators on some multiple time contour. For
two operators (shot noise) the standard 2-time Keldysh contour is
sufficient. However, to arrange more then two current operators
additional time axis must be included. The 3rd order correlation
function $S_3$ is describes by 4 lines contour (see Fig.1).
\cite{kogan}. The action consists of the integration over this
four lines path. However, as it is common for Keldysh technique,
we use only one infinite time path, though, with a four
independent field operators, corresponding to the  operators on
each four branches ( Fig.1) of complete time contour. Thus we
have:
\begin{equation}\label{action}
  S=\frac{i}{\hbar}\int_{-\infty}^{\infty}dt[\sum_{k\alpha\sigma}\hat{c}^{\dagger}_{ \alpha\sigma,k}
  \hat{G}^{-1}_{k\alpha} \hat{c}_{\alpha\sigma,k}-\sum_{j=1}^{4}\hat{\sigma}_{z}^{j}\hat{H}^{j}_{J}]
\end{equation}
Here $\hat{c}^{\dagger}_{\alpha\sigma,k},
\hat{c}_{\alpha\sigma,k}$ are
 the operators in $4\otimes4$ Keldysh space,
$\hat{G}_{k\alpha}$ denotes $4\otimes4$ Green's function of
noninteracting leads;
 $\sigma^{j}_{z}=1$ if j=1,3 and $\sigma^{j}_{z}=-1$ if j=2,4.
The interaction (\ref{H2}) now acquires Keldysh index $j$
$\rightarrow$$\hat{H}^{j}_{J}$. We use the important equalities
that relate the new Green functions on the contour in Fig1. to the
ordinary Keldish Green functions defined on two lines contour.
Keeping explicitly only Keldish indices we can write them as:
\begin{eqnarray}
G^{33}&=&G^{11},\, \,\ G^{44}=G^{22}\label{gk}\\
G^{jj'}&=& G^{12},\,\,\ j<j' \nonumber\\
       &=& G^{21},\,\,\ j>j' \nonumber
 \end{eqnarray}
 On this four fold contour the
correlation function is represented by time ordered product of
three current operators
\begin{eqnarray}
    {\cal S}_{3}(t_1 -t_2,t_2 -t_3)&=&
    \frac{1}{6}\sum_{P(ijk)}<TI^{i}(t_1 )I^{j}(t_2)I^{k}(t_3 )>
    \nonumber
\end{eqnarray}
Here superscripts $i,j,k$ are corresponding  Keldysh indices of
the Heisenberg operators (electrons and spin $\vec{S}$) related to
the first three paths in Fig.1 (each index runs from 1 to 3).
Summation  over all permutations $P$ of these indices is
performed. Now we are ready to use non-equilibrium Keldysh
technique and directly calculate $S_3$. In the lowest (fourth)
non-vanishing order of perturbation theory in the coupling
constant $J_{\alpha,\alpha'}$ the zero-frequency irreducible
correlation function $S_{3}$ acquires a form: $
S_{3}=S_{3}^{pot}+S_{3}^{pe}+S_{3}^{ex}$, with
$S_{3}^{pot}=Y_{-1}/( 2\hbar)$, $S_{3}^{pe}=9Y_{-1}/\hbar $ and
$S_{3}^{ex}=3Y_{7}/(2\hbar)$ where
\begin{eqnarray}
    Y_{n}&=&\frac{\pi^3 e^3g_{LR}^4}{4}\int
    d\omega(f_{L}-f_{R})(f_{L}(1-f_{R})+\nonumber\\
    &&f_{R}(1-f_{L})+
    n f_{L}f_{R})
\end{eqnarray}
Here $f_{L,R}$ are  Fermi distribution functions for leads, and
$g_{\alpha\alpha'}=J_{\alpha\alpha'}\nu$ stand for dimensionless
coupling constants. To mark
 the origin of different contributions we intentionally separate $S_{3}$
 into three parts: the first one $S_{3}^{pot}$ represents
 potential  scattering, the second part $S_{3}^{pe}$ defines mixed exchange and
potential interactions, while the last term $ S_{3}^{ex}$ stands
for pure exchange scattering processes. For temperatures
$T>>T_{K}$ ($T_{K}\simeq D_{0}\exp[-1/(g_{LL}+g_{RR})]$, $D_{0}$
are the Kondo temperature and the effective bandwidth,
correspondingly) logarithmic corrections which  appear in
perturbation
 theory can be disregarded. In this case the  above formulaes
 describe $S_{3}$ as a function of source-drain voltage (see inset
 in Fig.2 where  \underline{\emph{S}}= $S_{3}e$/$ T \sigma_{B}^2
 $ and $\sigma_{B}$= $\pi
e^2g_{LR} ^2 /\hbar $ is the conductance in the Born
approximation) for arbitrary values of the applied bias. It is
interesting to note that unlike the case of noninteracting system
a weak nonlinear voltage dependence of 3rd current cumulant at
$T>>T_K$ persists in the lowest order of perturbation theory.

 The logarithmic divergences appear in the next (fifth) order
expansion in couplings $g_{\alpha\alpha'}$. We compute $S_{3}$ to
this order only for $T=0$. In this case if voltages $eV>>T_{K}$
the perturbation theory still can be used. We keep only those
fifth order terms that consist of the maximal logarithmic
divergence. Only parts of $S_{3}$ which include exchange
tunnelling are affected by these logarithmic corrections, while
the pure potential part of the three currents correlation function
does not sufficiently changed. Thus after long, though, direct
calculations we arrive at
\begin{eqnarray}
   S_{3}^{pe}&=&\frac{9\pi^3}{4}\frac{e^{4}V}{\hbar}g_{LR}^4
   [1+2(g_{LL}+g_{RR})\ln\frac{D_{0}}{eV}]   \label{pe0}\\
   S_{3}^{ex}&=&\frac{3\pi^3}{8}\frac{e^{4}V}{\hbar}g_{LR}^4
   [1+4(g_{LL}+g_{RR})\ln\frac{D_{0}}{eV}]  \label{ex0}
\end{eqnarray}
When the voltage is decreasing  the log terms are starting to
increase so that for $eV \geq T_{K}$ the expansions
(\ref{pe0},\ref{ex0}) up to fifth order are not efficient. For
this region 3rd momentum of the noise can be derived in the
leading logarithmic approximation which consists of  summation of
the most diverging terms in each order in coupling constants $
g_{\alpha \alpha'}$. From a set of equations (\ref{pe0},\ref{ex0})
it becomes clear that the scaling behavior of $S_{3}$ is similar
to that of conductance \cite{glazman}. This becomes particulary
clear when we rewrite the expression for the 3rd noise in a
following form:
\begin{equation}\label{sf}
    S_{3}(T=0)=\pi V[\frac{1}{8}\sigma_{B}^2
    +3\sigma_{B} \sigma_{0}(V)+ \frac{2}{3}\sigma^{2}_{0}(V) ]
\end{equation}
where $\sigma_{0}(V)=
 3 \pi e^2 [1+2(g_{LL}+g_{RR})\ln(D_{0}/eV)]/(4 \hbar) $ is
 the 'spin exchange' part of conductance, the
part which includes logarithmic term. On the level of 'poor man'fs
scaling technique in the zero temperature limit the
renormalization proceeds till the band width $D$ becomes equal to
the applied bias $eV$. After that, the conductance can be
calculated in the Born approximation with the renormalized
exchange constant \cite{glazman}: $\sigma(V)=3\pi e^2
\Gamma_L\Gamma_R/ [2\sqrt{\hbar}(\Gamma_L+\Gamma_R)\ln
(eV/T_K)]^2$. The potential scattering contribution to conductance
$\sigma_{B}$ is not changed under RG transformations  and renders
a small correction to $S_{3}$. Thus, the final expression for the
3rd cumulant is given by equation (\ref{sf}) where we should
replace  $\sigma_{0}(V)$ on $\sigma(V)$.

{\it Fermi liquid regime}: To study regime where $T,eV <T_{K}$ we
apply the mean field slave boson approximation (MFSB) \cite{hew}.
The current operator for Anderson model acquires a form
\begin{equation}\label{cura}
    I_{s}=\frac{ie}{2\hbar}\sum\int_{\sigma}[(v_{L}
c^{\dagger}_{L \sigma}(0)-v_{R} c^{\dagger}_{R
\sigma}(0))d_{\sigma}(0)-H.C.]
\end{equation}
In the slave-boson approach, the localized electron operator
$d^{\dagger}_{\sigma}$ is represented by
$\hat{f}^\dagger_{\sigma}\hat{b}$ with $\hat{b}$ and
$\hat{f}^\dagger_{\sigma}$ being the standard boson and fermion
operators. The total action on the time contour in Fig.1 has the
similar form as (\ref{action}). We should replace only $H_{J}$ by
interacting $d$-dependent part of the Anderson Hamiltonian
(\ref{HA}). The two new terms in  the action, equivalent to the
first one in equation (\ref{action}) come from slave bosons  and
auxiliary fermions Green's functions. Also the requirement of a
single occupancy $\hat{b}^\dagger\hat{b}+
\sum_{\sigma}\hat{f}^\dagger_{\sigma}\hat{f}_{\sigma}=1$ must be
included into the action with a Lagrange multiplier.
  In the MFSB approximation  the Bose
 operators $\hat{b}^\dagger, \hat{b}$  are replaced by their expectation value
 $b$. We also add to the action the source term $S_{\gamma}=\int dt \gamma^j (t) I_{s}^j$
 (j=is running from 1 to 4) and integrate out the fermion operators in the
 leads. Here $I_{s}^j$ is the current operator for Anderson model
 taken
 on $j$ branch of time contour in Fig.1.
 For a symmetric tunnelling ( $v_L =v_R=v$, $\Gamma_{L}=\Gamma_{R}=\Gamma$ )
 the effective action can be written as  $S_{eff}(\gamma)=S_b+S_f$
 where the first term is nonoperator bosonic part of the action, while the last one explicitly depends
 on source fields $\gamma^{k}(t)$ and is given by
 \begin{eqnarray}\label{sof}
    S_{f}(\gamma)&=&\sum_{\sigma}\int dt \int dt'\hat{f}^\dagger_{\sigma}[G^{-1}_{d\sigma}(\gamma)]\hat{f}_{\sigma} \nonumber\\
    G^{-1}_{f\sigma}(\gamma)&=&G^{-1}_{0f\sigma}-T_{k}(Q^+G_{L\sigma}Q^- +
    Q^- G_{R\sigma}Q^{+}) \\
G^{-1}_{0f\sigma}(\omega)&=&(\omega-\tilde{\epsilon})\hat{\sigma_z},\,\,\
Q^{\pm}_{k}=\hat{\sigma}_{z}^k \pm \frac{ie}{2 \hbar}
\gamma^{k}(t) \nonumber
\end{eqnarray}
 Here $G_{L,R}$ represents $4\otimes4$ matrix of the electrons
propagators in the leads, $G_{0f\sigma}(\omega)$ is the Fourier
transform of zero order slave fermions Green's function and
$\hat{\sigma_z}$ is diagonal $4\otimes4$ matrix with elements
$\hat{\sigma_z}^j $ (\ref{action}). We also define $T_{k}=\Gamma
b^2$ as an effective Kondo temperature. The $G_{f\sigma}(\omega)$
includes the Lagrange multiplier which shifts the localized level
position $\epsilon$ to $\tilde{\epsilon}=\epsilon+\lambda $. Both
free parameters
 $T_{k}$ and renormalized level $\tilde{\epsilon}$ are self-consistently
  determined by two equations that define the extremum of
  $S_{eff}$ relative to $b$ and $\tilde{\epsilon} $ when
   $U\rightarrow\infty$  \cite{hew}.

 We can trace out
 the slave fermions and get a closed form for
generating functional:
$Z=exp[-\frac{i}{2\hbar}TrlnG^{-1}_{f}(\gamma)]$. The irreducible
three currents correlation function or the 3rd cunmulant is
obtained by taking the third variation of the logarithm of this
functional on source fields. Performing the straightforward
calculations we find $S_3 =S_3'+S_3''$ where
\begin{eqnarray}
  S_3'&=&\frac{\pi e^3}{2}\int
  d\omega \rho^2(\omega)(f_{L}-f_{R})^3(3-4\pi T_{k}
  \rho(\omega))\nonumber\\
  S_3''&=&\frac{\pi e^3}{2}\int
  d\omega \rho^2(\omega)(f_{L}-f_{R})[1-3(f_{L} + f_{R}-1)^2]\nonumber
\end{eqnarray}
Here the spectral density of interacting level
$\pi\rho(\omega)\equiv -Im
  G_{d}^r(\omega)=T_{k}/[(\omega-\tilde{\epsilon})^2 +T_{k}^2]$ is introduced.
If $T\rightarrow 0$ then 3rd cumulant becomes particulary simple.
Expression for $S_{3}$ is now reduced to
\begin{equation}\label{stem0}
  S_3=\frac{2}{\pi} e^3\int_{0}^{eV}d\omega
  T^2(\omega)(1-T(\omega))
\end{equation}
where $T(\omega)$= $\pi T_{k}\rho(\omega)$.
 Thus, we obtain 3rd noise $S_3$ as a function of two
dimensionless parameters $eV/T_{k}$ and $\tilde{\epsilon}/T_{k}$.
 A simple approximate
expression for $S_3$ follows when these parameters are small $
S_3\approx 2e^6V^3 /(3\pi T_{k}^2)$. Such a voltage dependence is
typical for a Fermi liquid.

{\it Discussion}: The crossover regime ($T, eV \sim T_{k}$)
actually is extended to a several threshold values $T_{K}$. This
region can also be studied within the present theory by using
non-crossing approximation in non-equilibrium (NCA)\cite{NCA}
together with $1/N$ expansion for generating functional $Z$. A
satisfied estimation of the 3rd noise in this case can be achieved
by calculating equation (\ref{stem0}) with a modified spectral
density which is derived in NCA approximation:
\begin{figure}[ht]
\begin{center}
\includegraphics [width=0.4\textwidth ]{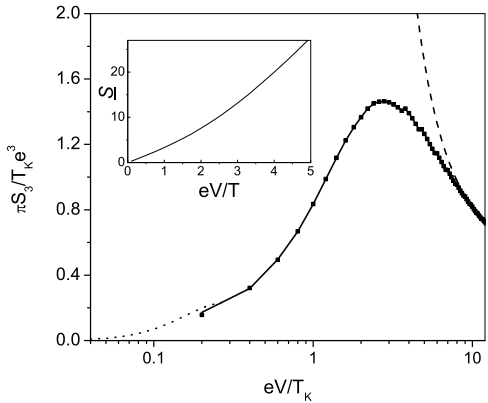}
\caption {\label{fig3} The third cumulant $S_{3}$ versus applied
voltage $ eV/ T_{K}$. Dot curve represents MFSB calculations valid
for small voltage $eV<T_{K}$. Solid line with black squares is
obtained in NCA approximation. Dash line is the result of direct
computation of Eq.(\ref{sf}) with $\sigma_0\rightarrow\sigma(V)$.
The curves which were calculated for two $T_{K}$ differing by a
factor of 10 practically coincide. This reflects the universality
of 3rd cumulant  in considered regime. ($\emph{inset}$):
Normalized 3rd cumulant $\underline{S}$ as function of $eV/T$ in
the high temperature limit $T>>T_{K} $.}
\end{center}
\end{figure}
$T_{K}\rho(\omega)\rightarrow\Gamma\rho_{NCA}(\omega)$. We start
numerical calculations by solving the mean field equations of MFSB
theory. The MFSB method  is limited to small voltages $eV<T_{K}$
where is known it gives the correct qualitative behavior to this
Fremi liquid regime. In addition, at $eV<<T_{K}$ the 3rd cumulant
in MFSB approximation  is matched with the perturbation theory
derivation at the unitary limit of Kondo Hamiltonian.  For
conductance and noise calculations \cite{us} the whole scheme with
MFSB works rather well. At $eV<T_{K}$
 MFSB also gives adequate description of 3rd cumulant which
is close to its exact value. The dot line in Fig.2 displayed the
results of our MFSB
 calculations.

  In the crossover region the NCA
 computations begin with a finite V.
The main block which we need to calculate in NCA is
$\rho(\omega)$. The 3rd current cumulant includes the frequency
integration of different powers of $\rho(\omega)$. To  estimate
the accuracy of NCA we consider the known  different physical
values which are based on spectral density derivation. For
example, for quantum dots, Cox \cite{cox} has shown that the
calculated equilibrium susceptibility agree with the exact Bethe
ansatz results to within the 0.5\% convergence accuracy of the
NCA. For the current calculations Wingreen and Meir (see
\cite{NCA}) use NCA equations in nonequilibrium and found
 at worse an overestimate of 15\% on the linear
response conductance. We expect the same (15\%) accuracy for 3rd
current cumulant. NCA result is given by solid curve with black
squares in Fig.2. Strong nonlinearity with a peak at the
intermediate voltage $eV_m \sim2.6T_{K}$ is determined by
significant Kondo correlations. Indeed, in the absence of
interacting $S_3$ grows linearly with $V$. At $eV\lesssim T_{K}$
we are in the
 strong Kondo regime where $S_3$ grows considerably
 faster then simply linearly  in $V$. The subsequent decreasing of 3rd noise
 for $V>V_m$ simply indicates on the weakening of the Kondo
 effect and manifests  that Kondo correlations continues to
 determine the physics 3rd noise. Even when we go
 outside the crossover region and enter scaling limit
 $D_{0}>>eV>>T_{K}$ (dash line on Fig.2)
 where Kondo model is in a weak coupling regime
 with depressed Kondo effect, $S_3$
 shows a behavior which is far for being linear in $V$. Though, the
  Kondo effect is depressed, nevertheless, the logarithmic contributions to 3rd
cumulant which in this region are related to Kondo correlations
are considerably larger than those derived with the help of bare
perturbation theory in the Born approximation. Linear or weak
nonlinear voltage dependence which would be indicate the complete
depression of Kondo correlations appears outside the domain of
Fig.2 for  bigger $T$ (as inset in Fig2) or $V$ ($T,V\sim D_{0}$).

 {\it Conclusions}: In conclusions, we have represented a general theory
 which allows to describe current fluctuations in quantum dot
 in the presence of interactions. In particular, we have considered a
 junction with quantum dot described by Kondo Hamiltonian. For this system we have
 calculated third cumulant of non-time
ordered product of three current operators. Restricting ourselves
to zero frequency we were able by our approach to cover
 three important regions: one is the in Fermi liquid limit, crossover
regime, and weak coupling Kondo limit. Unlike the non-interaction
system, in all cases $S_3$ shows nonlinear voltage dependence. The
linear dependence on $V$ is restored only for large voltages
$v\rightarrow \infty$. In this work we assumed that external
impedance is equal to zero, the condition, which may be violated
in the real experiment. This impedance leads to a modification of
the third cumulant \cite{nb}.
%---------------------------end read-------------------------------
\begin{acknowledgments}
 I would like to thank T. Martin for very helpful discussions
 and for drawing attention to the important work \cite{kogan}
and Yu. Nazarov  for his suggestions related to the theory of full
counting statistic. I also thankful to T. Aono, B. Horovitz, and
Y. Meir for numerous discussions.
\end{acknowledgments}


\begin{thebibliography}{99}
\bibitem{but} Ya.M. Blanter and M. B¨uttiker, Phys. Rep. {\bf 336}, 1 (2000).
\bibitem{reznikov} L. S. Levitov and M. Reznikov, cond-mat/0111057.
\bibitem{reulet} B.
Reulet, J. Senzier, and D.E. Prober, Phys. Rev. Lett. {\bf 91},
196601 (2003).
\bibitem{levitov} L.S. Levitov, H.W. Lee, and G.B. Lesovik, J. of
Math. Phys. 37, 4845 (1996).
\bibitem{nazarov}M. Kindermann and Yu.V. Nazarov,
Phys. Rev. Lett. 91, 136802 (2003).
\bibitem{nagaev}K.E. Nagaev, Phys. Rev. B {\bf66}, 075334 (2002).
\bibitem{gefen} D.B. Gutman and Yu. Gefen, Phys. Rev.
B {\bf68}, 035302 (2003).
\bibitem{zaikin} A.V. Galaktionov, D.S. Golubev, and A.D. Zaikin,
cond-mat/0403464; Phys. Rev. B \textbf{68}, 235333 (2003).
\bibitem{kogan} S.M. Kogan, Phys. Rev. A {\bf44}, 8072 (1991).
\bibitem{glazman} A. Kaminski, Yu.V. Nazarov, L.I. Glazman, Phys. Rev. B {\bf62},
8154 (2000).
\bibitem{ygal} Y. Meir, N. S. Wingreen, and P.A. Lee,
Phys. Rev. Lett. 70, 2601 (1993).
\bibitem{us} Y. Meir, A. Golub,
Phys. Rev. Lett. 88, 116802 (2002).
\bibitem{hew} A.C. Hewson, The Kondo Problem to Heavy Fermions
(Cambridge University Press, 1993).
\bibitem{NCA}
M. H. Hettler, J. Kroha and S. Hershfield, Phys. Rev. Lett. 73,
1967 (1994); Phys. Rev. B 58, 5649 (1998); N. Wingreen and Y.
Meir, Phys. Rev. B 49, 11040 (1994); P. Nordlander et al., Phys.
Rev. Lett. 83, 808 (1999); M. Plihal, D. C. Langreth, P.
Nordlander, Phys. Rev. B 61, R13341 (2000)., A. Rosch, et al.,
Phys. Rev. Lett. 87, 156802 (2001).
\bibitem{cox} Cox D. L.
Phys. Rev.  B35, 4561 (1987).
\bibitem{nb} C. W. J. Beenakker, M. Kindermann, and Yu. V. Nazarov
Phys. Rev. Lett. 90, 176802 (2003).

\end{thebibliography}
\end{document}